\documentclass[a4paper,12pt]{article}

\usepackage{amsfonts}
\usepackage{graphicx}
\usepackage[english,activeacute]{babel}
\usepackage{amsfonts,amssymb,amsthm,amsmath}
\usepackage{graphicx,epsfig}
\usepackage{natbib}
\usepackage{color}

\newcommand{\enegrita}{\mathbf{e}}
\newcommand{\phinegrita}{\mbox{\boldmath$\phi$}}
\newcommand{\bfi}{\mbox{\boldmath$\phi$}}

\newcommand{\pinegrita}{\mbox{\boldmath$\pi$}}

\newcommand{\blambda}{\mbox{\boldmath $\lambda$}}

\newcommand{\tetanegrita}{\mbox{\boldmath$\theta$}}

\newcommand{\be}{\begin{eqnarray}}
\newcommand{\ee}{\end{eqnarray}}
\newcommand{\beq}{\begin{equation}}
\newcommand{\eeq}{\end{equation}}
\newcommand{\ea}{\end{eqnarray*}}
\newcommand{\ba}{\begin{eqnarray*}}
\newcommand{\edi}{\end{displaymath}}
\newcommand{\bdi}{\begin{displaymath}}

\newcommand{\bt}{\mathbf{t}}
\newcommand{\hD}{\hat{D}}

\newcommand{\map}{\emph{MAP}}

\newcommand{\Qed}{\ensuremath{\Box}}

\begin{document}




\title{Maximum likelihood estimation in the two-state Markovian arrival process} 

\author{Emilio Carrizosa$^a$ and Pepa Ram\'irez-Cobo$^b$\\ 
{\small $^a$\emph{Departamento de Estad\'istica e Investigaci\'on Operativa, Universidad de Sevilla (Spain)},}\\ {\small $^b$ \emph{IMUS Instituto de Matem\'aticas de la Universidad de Sevilla (Spain)}}
}



\maketitle

\begin{abstract}
The Markovian arrival process ($\map$) has proven a versatile model
for fitting dependent and non-exponential interarrival times, with a
number of applications to queueing, teletraffic, reliability or
finance. Despite theoretical properties of $\map$s and models
involving $\map$s are well studied, their estimation remains less
explored. This paper examines maximum likelihood estimation of the
second-order $\map$  using a recently obtained parameterization of
the two-state $\map$s.
\end{abstract}



\section{Introduction}\label{sec: intro}
\label{sec: intro}
Since \cite{Neuts79} described the Markovian arrival processes ($\map$s for short) for the first time, a number of works have dealt with
theoretical properties and applications of such point processes.
In particular, because of their versatility, many uses in queueing, teletraffic, reliability or finance have been suggested.
For a recent account of the literature on $\map$s applications, we refer the reader to \cite{Kim,Wu,Okamura,Casale,Ocon2,Badescu,Cheung}.

The versatile character of $\map$s is due to two main properties; on
the one hand, the interarrival times (i.e, the times between epochs
of occurrence of a certain event) in a $\map$
have a phase-type distribution, which is a rather convenient and
flexible framework for fitting realworld data, see for example
\cite{Cinneide,Aalen,Asmussen3}. On the other hand, the $\map$ allows for correlated interarrival times,
a feature increasingly present in a number of real data traces.

While performance analysis for models incorporating $\map$s
is a well-developed area, less progress has been made on statistical
estimation for such models. The $\map$ is a complex model which
includes  transitions to hidden states between real arrivals. In
practice, only inter-arrival time data are usually observed and
therefore,  in this context, the observed data can be viewed as
being generated from a hidden Markov process. See e.g.\
\cite{Ephraim02}.

The simplest $\map$ is the two-state $\map$, called hereafter $\map_2.$ The $\map_2$ is usually represented in terms of six parameters, see
for instance \cite{Eum}, \cite{Bodrog} or \cite{Ramirez3}. However,
such representation in terms of $6$ parameters overparameterizes the
process, making it \emph{unidentifiable}: different $\map_{2}$
parameterizations produce the very same joint density for any
sequence of inter-arrival times, \citep{Ramirez}. In the context of
statistical inference, this implies that it is not sensible to
estimate the individual parameters of the $\map_2$ given a sample of
inter-arrival time data, since different parameters represent the
same process. Several papers have investigated a moments matching
approach for parameters inference, as is the case of \cite{Horvath},
\cite{Telek}, \cite{Eum}, \cite{Bodrog} or \cite{Casale}. However,
in these references, the issue of identifiability of the model has
not been taken into account (being \cite{Telek} and \cite{Bodrog} an
exception). Maximum likelihood estimation has been proposed
 in \cite{Breuer}, \cite{Klemm} and \cite{Okamura}, the EM algorithm being the tool suggested in such papers.
 The nonidentifiability of the representation used in terms of $6$ parameters has serious negative consequences:
 the likelihood function has infinitely many global maxima, and, on top of this,
the likelihood function may be highly multimodal, implying that standard methods such as the suggested EM algorithm,
will be strongly dependent on the starting values for these algorithms, and they run the risk of getting stuck at a poor local maximum.

Recently \cite{Bodrog} solve the identifiability problem for the
$\map_2$ by providing a canonical/unique representation of the
process, so that the infinitely many equivalent parameterizations
are reduced to a single one.

This work is intended as an attempt to gain
insight into the maximum likelihood estimation of the $\map_2.$ Unlike previous
studies, we do not use the EM algorithm, which calls for very time-consuming simulations in the "E" phase.
Instead, our analysis is based on the direct maximization of the likelihood
function.

This paper is organized as follows.
After a brief review of the second-order $\map$ in Section \ref{sec: map2}, we discuss
in Section \ref{sec: estimators}
how to compare estimators in the $\map_2.$ Then we describe in Section \ref{sec: MLE} the optimization problem consisting of maximizing the likelihood function.
Such maximization is not trivial, since technical problems appear for evaluating the objective and, needless to say, to optimize it.
The encountered numerical difficulties and the way to avoid them are pointed out in detail, and numerical illustrations are shown.

Finally, Section \ref{sec: discussion} discusses the findings and delineate some possible directions for future research.

\section{Preliminaries on $\map_2$s}\label{sec: map2}
The $\map_2$ is a doubly stochastic process $\{J(t),N(t)\}$, where $J(t)$ represents an irreducible, continuous,
 Markov process with state space $\mathcal{S}=\{1,2\}$ 
 and $N(t)$ is a counting process.  See
\citep{Neuts79,Lucantoni90,Lucantoni,Ramirez,Ramirez3}.

The $\map_2$ behaves as follows: the initial state $i_0\in \mathcal{S}$ is generated according to the initial probability vector
$\tetanegrita=(\theta,1-\theta)$ and at the end of an
exponentially distributed sojourn time in state $i$,  with mean
$1/\lambda_i$, two possible state transitions can occur. First,
with probability $0\leq p_{ij1}\leq 1$ a single arrival occurs and
the $\map_2$  enters a state $j\in \mathcal{S}$, which may be the same
as ($j=i$) or different to ($j \neq i$) the previous state.
On the other hand, with probability $0\leq p_{ij0}\leq 1$, no arrival occurs
and the $\map_{2}$ enters a different state $j \neq i$.

A stationary $\map_2$ can thus be expressed in terms of the parameters  $\{\blambda, P_{0}, P_{1}\},$  where $\blambda = (\lambda_{1},\lambda_{2})$, and $P_{0}$ and $P_{1}$ are $2\times 2$ transition probability matrices with elements $p_{ij0}$ ($i\neq j$) and $p_{ij1}$, respectively. Instead of transition probability matrices, any $\map_2$ can also be characterized by $\{D_{0},D_{1}\}$, in terms of the rate matrices,
\begin{equation}\label{known map}
D_{0}=\begin{pmatrix}
  -\lambda_1 &\ \lambda_1p_{120}  \\
  \lambda_2 p_{210} &\ -\lambda_2 \\
\end{pmatrix},\quad
D_{1}=\begin{pmatrix}
  \lambda_1p_{111} &\ \lambda_1(1-p_{120}-p_{111})  \\
  \lambda_2p_{211} &\ \lambda_2(1-p_{210}-p_{211}) \\
\end{pmatrix}.
\end{equation}
The matrix
$D_{0}$ is assumed to be stable, and as a consequence, it is nonsingular and the sojourn times are finite with probability 1. The definition of $D_{0}$ and $D_{1}$ implies that $D = D_0+D_1$ is the infinitesimal generator of the underlying Markov process, with stationary probability vector $\pinegrita=(\pi,1-\pi)$, computed as $\pinegrita D = {\bf 0}$.

The $\map_2$ can be viewed as a Markov renewal process. Indeed, let
$X_{n}$ denote the state of the  $\map_2$ at the time of the $n$th
arrival, and let $T_{n}$ denote the time between the $(n-1)$st and $n$th
arrival. Then $\{X_{n-1},T_{n}\}_{n=1}^\infty$ is a Markov renewal
process, and in particular,  $\{X_{n}\}_{n=1}^\infty$ is a Markov
chain whose transition matrix $P^\star$ is given by
\begin{equation}\label{Pstar}
P^\star =(-D_{0})^{-1}D_{1}.
\end{equation}

In practice
only partial information of the 
$\map_2$ is observed. It is 
assumed that the sequence of interarrival times $\{T_{n}\}_{n=1}^\infty$ is observed,
but the states where arrivals occur $\{X_{n}\}_{n=1}^\infty$ are not.

Special attention deserves the analysis of the random variable $T$, the time
between two successive arrivals in the stationary version of a $\map_2.$ Its moments are computed as
\begin{equation}\label{moments}
\mu_n=E(T^n)=n!\phinegrita \left(-D_0\right)^{-n} \enegrita,
\end{equation}
where $\phinegrita=(\phi, 1-\phi)$ is the probability distribution satisfying $\phinegrita P^\star = \phinegrita,$
and $\enegrita$ is a vector with all its coordinates equal to one.

The likelihood function for a sequence of interarrival times in the stationary version of the $\map_2$ is given by
\begin{equation}\label{lik}
f(t_1,t_2,\ldots,t_n|D_0,D_1) =\bfi e^{D_{0}t_{1}}D_{1} e^{D_{0}t_{2}}D_{1}\ldots e^{D_{0}t_{n}}D_{1}  \enegrita.
\end{equation}

Observe that the $\map$ allows for correlated inter-arrival times, thus the likelihood
function in (\ref{lik}) does not decompose into the product of the marginal likelihoods of the different terms.
The coefficient $\rho_k$ of autocorrelation of lag $k$ is given by
\begin{equation}\label{acf_map2}
\rho_k=\gamma^{k}\ \frac{\displaystyle\frac{\mu_2}{2}-\mu_1^2}{\mu_2-\mu_1^2}, \quad \text{for } k>0,
\end{equation}
where $0\leq \gamma<1$ is one of the two eigenvalues of the transition matrix $P^\star$
(since $P^\star$ is stochastic, then necessarily the other eigenvalue is equal to 1), \citep{Bodrog}.

\par\medskip
The expression (\ref{known map}) for the $\map_2$ in terms of $6$ parameters is known to be overparameterized, \cite{Ramirez}.
However, \cite{Bodrog} provide a unique, canonical representation for the $\map_{2}$ in terms of just four parameters.
Such canonical representation is the one we are using in this paper.
Specifically, if the correlation parameter $\gamma$ in (\ref{acf_map2}) is positive, then the canonical form of the $\map_2$ is given by
\begin{equation}\label{can1}
D_{0}=\left(\begin{array}{cc} x & y\\
0 & u\end{array}\right), \qquad D_{1}=\left(\begin{array}{cc} -x-y & 0  \\
v & -u-v \end{array}\right).
\end{equation}
On the other hand, for those $\map_2$s such that $\gamma\leq 0$, then their canonical form is
\begin{equation}\label{can2}
D_{0}=\left(\begin{array}{cc} x &  y\\
0 & u\end{array}\right), \qquad D_{1}=\left(\begin{array}{cc} 0 & -x-y  \\
-u-v & v \end{array}\right),
\end{equation}
where, $x,u\leq 0$, $y,v \geq 0$, $x+y \leq 0$, $u+v\leq 0$.

\section{Comparing estimators}\label{sec: estimators}
Our aim is to derive (maximum likelihood) estimates of the parameters of the $\map_2$s. This would allow one, for instance,
to make inference on the distribution function
of the random variable $T,$ or to properly simulate the process.

A remarkable issue is that  $\map_2$s may have very \emph{similar}
behavior, despite being represented by rather different parameters. This is notable since traditionally the $\map_2$ has been analyzed using
the overparameterized form (\ref{known map}): pretty different parameters sets are fully equivalent, in the sense that they
represent exactly the same $\map_2.$ Even if the canonical form (\ref{can1})-(\ref{can2}) is used,
and thus no indentifiability problems exist,
different parameters may yield very similar $\map$s. In other words,
closeness of two $\map_2$s 
is not correctly measured in terms of the (euclidean) distance
between the parameters identifying them.  In order to adequately
compare different estimators we may use as similarity measure
between them a similarity measure of the processes they represent.
In particular, we measure closeness between parameters representing
two $\map_2$s by an empirical Kullback-Leibler divergence (from now
on KL 
divergence) of their interarrival times joint density functions:
Given two $\map_2$s, with associated matrices $\{D_0,D_1\}$ and
$\{\hD_0,\hD_1\}$, given the length  $n$ of the observed sequences
and the number $N$ of runs the experiment is repeated, we will
measure the closeness between two $\map_2$s by means of the
empirical KL divergence
$D_{KL}\left(\{D_0,D_1\}||\{\hD_0,\hD_1\}\right),$
\begin{equation*}
D_{KL}\left(\{D_0,D_1\}||\{\hD_0,\hD_1\}\right):= \frac{1}{N}\sum_{i=1}^N\log \frac{f({\bf t}^{(i)}|\{D_0,D_1\})}{f({\bf t^{(i)}}|\{\hD_0,\hD_1\})},
\end{equation*}
where, for $i=1,2,\ldots,N,$ ${\bf t}^{(i)}=(t^{(i)}_1,\ldots,t^{(i)}_n)$ is a sequence of interarrival times generated from $\{D_0,D_1\}.$

\vspace*{0.1in} \noindent{\it Example 1.~} As an example, we consider a sample of $n=500$ interarrival times simulated from the $\map_2$
with canonical form
\begin{equation}\label{ex1}
D_{0}=\left(\begin{array}{cc} -20 & 6\\
0 & -0.5\end{array}\right), \qquad D_{1}=\left(\begin{array}{cc} 14 & 0  \\
0.0426 & 0.4574 \end{array}\right).
\end{equation}
We want to compare estimates as obtained from the method of moments, as discussed in Section \ref{mm} below.
The theoretical and empirical moments are given respectively by
\begin{eqnarray}\nonumber
(\rho_1,\mu_1,\mu_2,\mu_3)&=&\left(0.0864,1.6802,6.6887,40.1276\right),\\ \label{emp_moments}
(\bar{\rho_1 },\bar{\mu}_1,\bar{\mu}_2,\bar{\mu}_3)&=&\left(0.0643,1.6494,7.0219,44.1291\right).
\end{eqnarray}

The estimate is given by
\begin{equation}\label{est1}
\hat{D}^{(1)}_0(0)=\left(\begin{array}{cc}-999.9998 & 500.5033\\
0 & -0.4735\end{array}\right), \qquad \hat{D}^{(1)}_1(0)=\left(\begin{array}{cc} 499.4965 & 0  \\
0.1315 & 0.3420 \end{array}\right),
\end{equation}
with moments given by
$$(\hat{\rho}_1,\hat{\mu}_1,\hat{\mu}_2,\hat{\mu}_3)=(0.0643,1.6538,6.9842,44.2471).$$
{The superscript $^{(1)}$ in (\ref{est1}) implies that the $\map_2$ is expressed in the first canonical form. On the other hand, the notation $\hat{D}^{(1)}_0(0)$ and $\hat{D}^{(1)}_1(0)$ in (\ref{est1}) refers to the initial solution to the ML problem (see Section \ref{sec: MLE}).} If instead, a sample of size $n=1000$ is considered, the empirical moments,
$$(\bar{\rho}_1,\bar{\mu}_1,\bar{\mu}_2,\bar{\mu}_3)=\left(0.0804,1.6877,6.7973,43.0030\right),$$
are closer to the theoretical ones, and the estimate is given by
\begin{equation*}
\hat{D}^{(1)}_0(0)=\left(\begin{array}{cc}-2.1562 &0.6346\\
0 & -0.4679\end{array}\right), \qquad \hat{D}^{(1)}_1(0)=\left(\begin{array}{cc} 1.5216 & 0  \\
0.0852 & 0.3827 \end{array}\right),
\end{equation*}
whose moments are
$$(\hat{\rho}_1,\hat{\mu}_1,\hat{\mu}_2,\hat{\mu}_3)=\left(0.0804,1.6877,6.7973,43.0034\right).$$

It is interesting to note that, despite the estimated moments are close to the empirical and theoretical values, the elements of the matrices $\{\hat{D}^{(1)}_0(0),\hat{D}^{(1)}_1(0)\}$ for $n=500$ and $n=1000$ differ pretty much from those of  the theoretical $\{D_0,D_1\}$ (with exception of parameter $u$). If the empirical moments in the objective function are replaced by the real, theoretical ones, then the estimated matrices become
\begin{equation*}
\hat{D}^{(1)}_0(0)=\left(\begin{array}{cc}-21.9163 &6.5879\\
0 & -0.5001\end{array}\right), \qquad \hat{D}^{(1)}_1(0)=\left(\begin{array}{cc} 15.3284 & 0  \\
0.0425 & 0.4576 \end{array}\right),
\end{equation*}
more similar to the theoretical $\{D_0,D_1\}$.

The empirical KL divergences give us a more informative image on how far the estimated processes are from the original one:
\begin{eqnarray}\nonumber
D_{KL}\left(\{D_0,D_1\},\{\hD^{(1)}_0(0),\hD^{(1)}_1(0)\}\right)&=&46.0405 \ (n=500),\\ \label{DKLs}
D_{KL}\left(\{D_0,D_1\},\{\hD^{(1)}_0(0),\hD^{(1)}_1(0)\}\right)&=&9.6855 \ (n=1000),\\ \nonumber
D_{KL}\left(\{D_0,D_1\},\{\hD^{(1)}_0(0),\hD^{(1)}_1(0)\}\right)&=& 0.0430 \ (\text{theoretical moments}).
\end{eqnarray}
From the above results, we can assert that estimate in the case where the empirical moments are exactly the theoretical ones is closer to $\{D_0,D_1\}$, than the estimate when $n=1000$, which is closer to $\{D_0,D_1\}$ than the estimate in the case that $n=500$. However, since the DK divergence is not upper bounded, the value $46.0405$ is not conclusive enough of how similar $\{D_0,D_1\}$ and its estimate are. In order to get a clearer idea of this,
a random different $\map_2$ from (\ref{ex1}) was simulated
\begin{equation}\label{ex2}
D^\star_0=\left(\begin{array}{cc}-1 & 0.001\\
0 & -0.005\end{array}\right), \qquad D^{\star}_1=\left(\begin{array}{cc} 0.999 & 0  \\
10^{-5} & -10^{-5}+0.005 \end{array}\right),
\end{equation}
with theoretical moments
\begin{equation}
\label{moments_ex2}
\left(\hat{\rho}(1),\hat{\mu}_1,\hat{\mu}_2,\hat{\mu}_3\right)=\left(0.3963,67.3783,2.6686\times 10^4,1.6011\times 10^7\right).
\end{equation}

Then, we obtained
\[D_{KL}\left(\{D_0,D_1\},\{D^{\star}_0,D^{\star}_1\}\right)=74.9794,\]
which is clearly larger than the divergences in (\ref{DKLs}).

Although the previous results are preliminary, they shed some light on the complexity when comparing two given $\map_2$ representations.
Since the topic exceeds the scope of this paper we do not look into it in greater depth and aim to address it in the future. \Qed

\section{Maximum likelihood estimate}\label{sec: MLE}
In this section we look closely at the problem of estimating the parameters in the $\map_2$ by maximizing the likelihood function, given by (\ref{lik}).
We will make use of the canonical representation of the process, and this way we avoid the typical switching problems of nonidentifiability.
Specifically,  given a sequence of interarrival times ${\bf t}=(t_1,t_2,\ldots,t_n)$ we aim
to solve the following optimization problem, concerning the first canonical form:
\begin{equation*}(P1)\label{problema}
\left\{
\begin{array}{lll}
\max & \bfi e^{D_{0}t_{1}}D_{1} e^{D_{0}t_{2}}D_{1}\ldots e^{D_{0}t_{n}}D_{1}\enegrita   \\
\\
\mbox{s.t.} &  D_{0}=\left(\begin{array}{cc} x & y\\
0 & u\end{array}\right), \\ \\
 & D_{1}=\left(\begin{array}{cc} -x-y & 0  \\
v & -u-v \end{array}\right),\\ \\
 &x,u\leq 0,\\
 &y,v \geq 0,\\
 &x+y\leq 0,\\
 &u+v\leq 0,\\
 & \phinegrita (-D_0)^{-1}D_1 = \phinegrita.
\end{array} \right.
\end{equation*}
With regard to the second canonical form, we formulate $(P2)$ as $(P1)$, where matrices $D_0$ and $D_1$ are given by (\ref{can2}). To obtain the $\map_2$ estimate, we proceed as follows.
First, the solutions to $(P1)$ and $(P2)$, $\{\hat{D}^{(1)}_{0},\hat{D}^{(1)}_{1}\}$ and $\{\hat{D}^{(2)}_{0},\hat{D}^{(2)}_{1}\}$ are computed.
Finally, the selected estimate will be the $\map_2$ $\{\hat{D}^{(1)}_{0},\hat{D}^{(1)}_{1}\}$ or $\{\hat{D}^{(2)}_{0},\hat{D}^{(2)}_{1}\}$
that maximizes the likelihood.

Textbook models usually simplify maximum likelihood estimation problems by taking logs, and then simplifying the objective,
which is given as a summation
of $n$ terms. This is not possible in our model: the objective function (\ref{lik}) does not admit such a factorization
due to the fact that the interarrival times are not independent, and thus the joint density is not expressed as the product of marginal likelihoods.
This makes even the evaluation of the objective cumbersome.
Other technical difficulties also appear. These, as well as ways to overcome such difficulties, are discussed in what follows.

\subsection{Finding a starting solution}\label{mm}
The choice of a good starting solution is always crucial to attain convergence of the ML algorithm to a good estimate. This is particularly relevant
in our case, since an inadequate choice of the parameters may lead the algorithm to diverge, or even to be unable to provide an output, because of the
presence of too big numbers.

We have found that a good starting point is obtained if one uses the moments matching estimate. The procedure to derive it is described below.
The canonical representation of the $\map_2$ in terms of four parameters leads \cite{Bodrog} to show that any $\map_2$ is completely characterized by
its first three moments, $\mu_1$, $\mu_2$, $\mu_3$ and lag-one autocorrelation coefficient $\rho_1$.
As a consequence,
given a sequence of interarrival times ${\bf t}=(t_1,t_2,\ldots,t_n)$ with sample values $\bar{\mu_i}$, for $i=1,2,3$ and $\bar{\rho}(1)$,
the method of moments
would allow one to estimate the parameters $(x,y,u,v)$ in the canonical form of the $\map_2$ by solving the nonlinear system of equations
\begin{equation}
\label{eq:mm_equation}
\begin{array}{rcl}
\mu_i(x,y,u,v) & = & \bar{\mu_i},\quad \text{for}\quad i=1,2,3,\\
\rho_1(x,y,u,v), & = & \bar{\rho}_1.  \end{array}
\end{equation}

However, in real-world data,  (\ref{eq:mm_equation}) may have no feasible solution. In order to obtain an estimate, we seek instead the parameters
$(x,y,u,v)$ fulfilling as much as possible (\ref{eq:mm_equation}). Given $\tau >0, $ define the function
\begin{eqnarray*}
\!\!\!\!\!\!\delta_\tau(x,y,u,v) &=& \left\{\rho_1(x,y,u,v)-\bar{\rho_1} \right\}^2 +\\
&+&
\tau \left\{\left(\frac{\mu_1(x,y,u,v)-\bar{\mu}_1}{\bar{\mu}_1}\right)^2   + \left(\frac{\mu_2(x,y,u,v)-\bar{\mu}_2}{\bar{\mu}_2}\right)^2 +\left(\frac{\mu_3(x,y,u,v)-\bar{\mu}_3}{\bar{\mu}_3}\right)^2  \right\}.
\end{eqnarray*}

We propose to solve the following optimization problem:
\begin{equation*}(P0)
\left\{
\begin{array}{lll}
\min & \displaystyle \delta_\tau(x,y,u,v) \\
\mbox{s.t.} & x,u\leq 0, \\
 & y,v\geq 0,\\
 &x+y\leq 0,\\
 &u+v\leq 0.
\end{array}
\right.
\end{equation*}

The penalty parameter $\tau$ needs to be tuned. In our experiments it has been set to $\tau=1$, which seems to perform well in practice.
Obviously $(x,y,u,v)$ solves (\ref{eq:mm_equation}) iff it is an optimal solution of (P0), whose optimal value is $0.$

In order to solve the multimodal Problem (P0), we have used the
 MATLAB${}^\copyright$ routine {\tt fmincon}. Numerical inaccuracies were found, and then the
range of the parameters was slightly reduced, by adding to (P0) the constraints
\begin{eqnarray*}
x,u & \in &
[-1000,-2 \times10^{-16}] \\
y,v & \in & [0.00001,100].
\end{eqnarray*}
A  multistart was then executed with $100$ randomly
chosen starting points and found to yield satisfactory
results. The solution to (P0), noted $\{\hat{D_0}(0),\hat{D_1}(0)\}$ will be used
as starting point of the algorithm that maximizes the likelihood function.

{It is worth pointing out here that other initial values could have been chosen, for example random starting $\map_2$s; however we have found that the use of the moments matching estimate reduces the numerical problems in practice. A total of one thousand random $\map_2$s were estimated via the ML method described in Section \ref{eval_lik} where the starting values were (1) randomly generated versus (2) the moments matching estimates.
In the first case,
in a $32\%$ of the generated $\map_2,$ the solution given by the
computer possessed a likelihood function equal to 0 or
to 
infinite. This percentage decreased to $14\%$ in the case of the
moments matching estimates. When the objective function was
evaluated using the final ML estimates, a $35\%$ of times it was
equal to 0 or to
infinite in the first case (that is, when a random seed was
selected), against a $1\%$ when the moments matching estimate was
used as starting value. Additionally, for those cases where the
objective function did not present any numerical inconsistency using
a random starting point, the $61.53\%$ of times the objective
function was larger using a moments method estimate as starting
point than when a random $\map_2$ was used.}

\subsection{Evaluation of the likelihood function}\label{eval_lik}
In principle, (P1)-(P2) can be solved using standard optimization routines, and, as discussed above, the moments method estimate, obtained solving (P0)
with a multistart, is a recommended starting point.

However we have found serious difficulties in carrying out the numerical evaluation of the likelihood function (\ref{lik}),
which turns out problematic in practice when the variability in the sample $\bt$ is {\it large}.
This section is devoted to analyze such a problem.

\begin{figure}[htb]
\begin{center}
\begin{tabular}{c}
\includegraphics[height=2.7in]{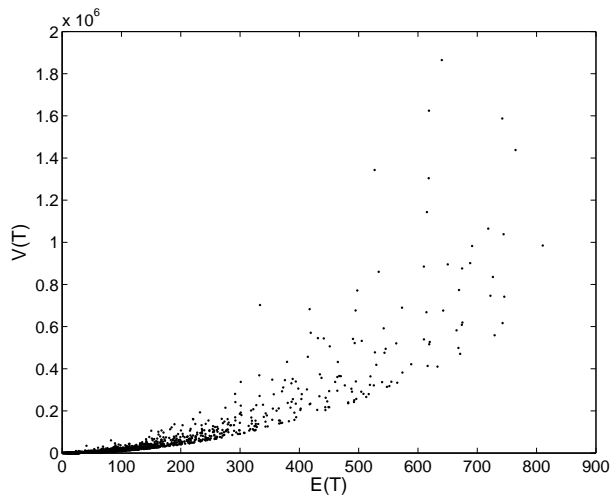}
\end{tabular}
\caption{\it $E(T)$ vs. $V(T)$ for 100000 simulated random MAP$_{2}$s.}\label{muvar}
\end{center}
\end{figure}

As a motivational example, consider the $\map_2$ given by (\ref{ex2}). Note that the theoretical variance of the interarrival times is $2.2146\times 10^4$. A sample of 500 observations was generated from this $\map_{2}$ with a sample variance equal to $3.4521\times 10^{4}$.  That is why some extreme values, of the order of $10^{3}$ were obtained. When evaluating the likelihood (\ref{lik}), it was found that $f(t_1,\ldots,t_n|D_0,D_1)\approx 0$. An explanation for this phenomenon is as follows. Given a $\map_2$ with canonical form as in (\ref{can1}),
the term $e^{D_{0}t}D_{1}$ in (\ref{lik}) satisfies
\[e^{D_{0}t}D_{1}=\left(\begin{array}{ccc} (-x-y)e^{tx}+y\displaystyle\frac{e^{tx}-e^{tu}}{(x-u)v} & & y\displaystyle\frac{e^{tx}-e^{tu}}{(x-u)(-u-v)}\\
& &\\
ve^{tu} & & (-u-v)e^{tu}\end{array}\right).\]
Since $x,u<0$, it follows immediately that
\[\lim_{t\rightarrow \infty}e^{D_{0}t}D_{1}=\mathbf{0},\]
no matter which values the parameters $(x,u,y,v)$ take.  Here $\mathbf{0}$ denotes a $2\times 2$ zero matrix.
The same phenomenon happens when the second canonical form is considered. This result implies that, in practice,
in the presence of \emph{large} interrarival times, the numerical evaluation of (\ref{lik}) is
rather difficult.
For instance, in the considered sample, $t_1 = 18.12$, $t_2=465.49$, $t_3=120.70$ and
\[e^{D_{0}t_1}D_{1}= \left(\begin{array}{cc}0.0000 & 0.0000\\
0.0000 & 0.0046\end{array}\right), \]
\[e^{D_{0}t_1}D_{1}e^{D_{0}t_2}D_{1}=10^{-5}\times \left(\begin{array}{cc}0.0000 & 0.0002\\
0.0004 & 0.2218\end{array}\right), \]
\[e^{D_{0}t_1}D_{1}e^{D_{0}t_2}D_{1}=10^{-8}\times \left(\begin{array}{cc}0.0000 & 0.0006\\
0.00012 & 0.6054\end{array}\right). \]
The factors $e^{D_{0}t_1}D_{1}\ldots e^{D_{0}t_k}D_{1}$ become smaller as $k$ increases, and indeed the computer (MATLAB${}^\copyright$ software) returns $e^{D_{0}t_1}D_{1}\ldots e^{D_{0}t_k}D_{1}={\bf 0}$ for $k = 118$. This example is not an isolated case. Indeed, we experienced that it is more a rule than an exception that large interarrival times appear in the simulated samples. From Figure \ref{muvar}, which depicts the theoretical variance versus the mean of the inter-arrival times of $100,000$ randomly simulated $\map_{2}$s, it can be seen that the variance $V(T)$ increases considerably with the mean $E(T)$.

We have found rather convenient to \emph{re-scale} the sample, thus re-scaling the likelihood function to a more tractable range. This is possible
since the likelihood function (\ref{lik}) satisfies
\begin{equation}\label{rescale}
f(\bt |D_0,D_1)=c^{-n}f\left(\frac{1}{c}\bt \ | \ cD_0,cD_1\right) \qquad \forall c > 0,
\end{equation}
where $n$ is the length of {\bf t}. In other words, the ML estimates obtained for interarrival times ${\bt}$ is a re-scaled by $c$ version of that obtained for
interarrival times $\displaystyle \frac{1}{c}{\bt}$, for any positive $c$. In our numerical experience we have found good results setting $c$ as the standard deviation of the data, so that the new sample variance is equal to $1$ and therefore, less extreme values are expected to appear in the sample.
Specifically, the algorithm to follow is:
{\tt
\begin{enumerate}
\item Set $c:=std({\bf t})$, the standard deviation of ${\bf t}=\left(t_1,\ldots,t_n\right).$
\item Consider the new sample ${\bt^\star}=\displaystyle\left(\frac{1}{c}\bt\right)$.
\item Compute the ML estimates of $D^{\star}_0$ and $D^{\star}_1$, noted $\hD^\star_0$ and $\hD^\star_1$,
 by maximizing $f(\bt^\star |D^{\star}_0,D^{\star}_1)$ via a standard optimization algorithm.
\item Calculate the estimate of $D_0$ and $D_1$ as $\hD_0=\displaystyle \frac{1}{c}\hD^\star_0$ and $\hD_1=\displaystyle \frac{1}{c}\hD^\star_1$.
\end{enumerate}
}

Next section illustrates the approach for a pair of simulated data sets.

\subsection{Numerical illustration}

\vspace*{0.1in} \noindent{\it Example 2.~}
Consider the sequence of interarrival times, the $\map_2$ defined by (\ref{ex1}) and its moments matching estimate (\ref{est1}) in {\it Example 1}.
It can be checked that $f\left({\bf t} |\hD^{(1)}_0(0),\hD^{(1)}_1(0)\right) \approx 0$. We set $c = std({\bf t})=2.076$ and we compute
${\bf t}^\star={\bf t}/c$. Now, it can be seen that
\[\log f\left({\bf t}^\star|c \hD^{(1)}_0(0),c \hD^{(1)}_1(0)  \right)=-431.3554.\] From (\ref{rescale}), it can be concluded that
\[\log f\left({\bf t}| \hD^{(1)}_0(0), \hD^{(1)}_1(0)  \right)=-500\times\log(c)-431.3554 = -796.5823.\]
The MATLAB$^\copyright$ routine {\tt fmincon} is used to obtain the solution to (P1), the ML estimates $\{\hD^\star_0,\hD^\star_1\}$. In this case, it was found that
\begin{equation*}
\hD^\star_0=\left(\begin{array}{cc}-30.7238 & 6.7257\\
0 & -1.0069\end{array}\right), \qquad \hD^\star_1=\left(\begin{array}{cc} 23.9981 & 0  \\
0.0735 & 0.9334 \end{array}\right),
\end{equation*}
and then, dividing $\hD^\star_0$ and $\hD^\star_1$ by $c$ yields
\begin{equation}\label{est4}
\hD^{(1)}_0=\left(\begin{array}{cc}-14.7994 & 3.2397\\
0 & -0.4850\end{array}\right), \qquad \hD^{(1)}_1=\left(\begin{array}{cc} 11.5596 & 0  \\
0.0354 & 0.4496 \end{array}\right).
\end{equation}
whose moments $\{\rho_1,\mu_1,\mu_2,\mu_3\}$ are obtained as
\[\left(\rho_1,\mu_1,\mu_2,\mu_3\right) = \left(0.1163,1.6537,6.7643,41.8285\right). \]
In addition, the log-likelihood has increased with respect to the one provided by the moments matching estimate, i.e., the one obtained by solving (P0):
\[\log f\left(\bt^\star|\hD^\star_0,\hD^\star_1  \right)=-248.5386,\] which implies that
\begin{equation}\label{loglik_ex2}
\log f\left(\bt| \hD^{(1)}_0, \hD^{(1)}_1  \right)=-613.7655.
\end{equation} There has been also an improvement in terms of the DK divergence:
\begin{equation}\label{DKL_ex2}
D_{KL}\left(\{D_0,D_1\},\{\hD^{(1)}_0,\hD^{(1)}_1\}\right)=0.7938,
\end{equation} considerably smaller than $46.0405$ in (\ref{DKLs}).

{Next, we consider the estimate of the $\map_2$ in the second canonical form. The solution to (P0) is found
\begin{equation*}
\hat{D}^{(2)}_0(0)=\left(\begin{array}{cc}-77.6722 & 23.4148\\
0 & -0.4861\end{array}\right), \qquad \hat{D}^{(2)}_1(0)=\left(\begin{array}{cc} 0 & 54.2573  \\
0.1406 & 0.3455 \end{array}\right),
\end{equation*}
with estimated moments
\begin{equation*}
\left(\hat{\rho}(1),\hat{\mu}_1,\hat{\mu}_2,\hat{\mu}_3\right)=\left(-0.0287,1.7144,7.0451,43.4798\right).
\end{equation*}
Note how the estimated moments are close to the empirical ones given by (\ref{emp_moments}), with the exception of the autocorrelation coefficient, which in this case is negative. The algorithm to solve (P2) was implemented with starting solution given by $\{\hD^{(2)}_0(0),\hD^{(2)}_1(0)\}$ and yielded
\begin{equation*}
\hD^{(2)}_0=\left(\begin{array}{cc}-16.8292 & 6.4688\\
0 & -0.5343\end{array}\right), \qquad \hD^{(2)}_1=\left(\begin{array}{cc} 0 & 10.3606  \\
0.1236 & 0.4107 \end{array}\right).
\end{equation*}
whose moments are
\[\left(\rho_1,\mu_1,\mu_2,\mu_3\right) = \left(-0.0146,1.6505,6.1523,34.5420\right). \] The KL divergence is
\[D_{KL}\left(\{D_0,D_1\},\{\hD^{(2)}_0,\hD^{(2)}_1\}\right)=10.0508,\] larger than (\ref{DKL_ex2}).  Finally, the log-likelihood function is
\begin{equation}\label{loglik_ex22}
\log f\left(\bt| \hD^{(2)}_0, \hD^{(2)}_1  \right)=-707.3972.
\end{equation}
To select the final estimate, the log-likelihoods (\ref{loglik_ex2}) and (\ref{loglik_ex22}) are compared. In this case the estimate of the $\map_2$ in its first form is chosen.} \Qed

\vspace*{0.1in} \noindent{\it Example 3.~}
In this example, a $\map_2$ with a large variance of the interarrival times is estimated.
Consider the $\map_2$ defined by (\ref{ex2}), whose theoretical moments are given by (\ref{moments_ex2}).
Note that the variance of the interarrival time is $2.2146\times 10^4$.
Let ${\bf t}= (t_1,\ldots,t_n)$ be a sample of size $n=500$ of interarrival times simulated from (\ref{ex2}) whose sample moments are
$$(\bar{\rho}_1,\bar{\mu}_1,\bar{\mu}_2,\bar{\mu}_3)=\left(0.0709,180.6187,6.9675\times 10^4, 3.8681\times 10^7\right).$$ In this case, $c=std({\bf t})=192.6803$.
The estimate obtained by solving (P0) in the first canonical form is given by
\begin{equation*}
\hat{D}^{(1)}_0(0)=\left(\begin{array}{cc}-122.6681 & 24.7206\\
0 & -0.0052\end{array}\right), \qquad \hat{D}^{(1)}_1(0)=\left(\begin{array}{cc} 97.9475 & 0  \\
0.0001 & 0.0051 \end{array}\right).
\end{equation*}

The moments of the $\map_2$ defined by $\{\hat{D}^{(1)}_0(0),\hat{D}^{(1)}_1(0)\}$ are
$$(\hat{\rho}_1,\hat{\mu}_1,\hat{\mu}_2,\hat{\mu}_3)=(0.0516,170.2164,6.8348\times 10^4,3.9318\times 10^7).$$
As in \emph{Example 2}, it can be checked that
$f\left({\bf t}|\hD^{(1)}_0(0),\hD^{(1)}_1(0)\right)\approx 0$. However,
\[\log f\left({\bf t}^\star|c \hD^{(1)}_0(0),c \hD^{(1)}_1(0)  \right)=-466.9192,\] which, from (\ref{rescale}), implies
\[\log f\left({\bf t}| \hD^{(1)}_0(0), \hD^{(1)}_1(0)  \right)= -3097.4.\] The solution to (P1) was
\begin{equation*}
\hD^\star_0=\left(\begin{array}{cc}-200.0788 & 1.3418\\
0 & -0.9944\end{array}\right), \qquad \hD^\star_1=\left(\begin{array}{cc} 198.7370 & 0  \\
0.0019 & 0.9925 \end{array}\right),
\end{equation*}
and then, dividing $\hD^\star_0$ and $\hD^\star_1$ by $c$ leads to
\begin{equation*}
\hD^{(1)}_0=\left(\begin{array}{cc}-1.0384 & 0.0070\\
0 & 0.0052\end{array}\right), \qquad \hD^{(1)}_1=\left(\begin{array}{cc} 1.0314& 0  \\
0.0000 & 0.0052\end{array}\right).
\end{equation*}
It can be seen that the log-likelihood function has increased to
\[\log f\left({\bf t}^\star|\hD^\star_0,\hD^\star_1  \right)=-392.8464,\] or equivalently,
\[\log f\left({\bf t}| \hD^{(1)}_0, \hD^{(1)}_1  \right)=-3023.4.\]

The estimated moments are
$$(\hat{\rho}_1,\hat{\mu}_1,\hat{\mu}_2,\hat{\mu}_3)=\left(0.1765,151.6006,5.8668\times 10^4,3.4103\times 10^7\right).$$

Finally, the DK divergence with respect to the estimates are
\begin{equation}\label{DKL_ex3}
D_{KL}\left(\{D_0,D_1\},\{\hD^{(1)}_0,\hD^{(1)}_1\}\right)=0.6973,\end{equation} much smaller than that obtained from (P0):
\begin{equation*}
D_{KL}\left(\{D_0,D_1\},\{\hD^{(1)}_0(0),\hD^{(1)}_1(0)\}\right)=151.6021.
\end{equation*}

{The solution to (P0), in second canonical form was found as
\begin{equation*}
\hat{D}^{(2)}_0(0)=\left(\begin{array}{cc}-0.0053 & 0.0053\\
0 & -142.8445\end{array}\right), \qquad \hat{D}^{(2)}_1(0)=\left(\begin{array}{cc} 0 & 0  \\
137.4806 & 5.3639 \end{array}\right),
\end{equation*}
with estimated moments
\begin{equation*}
\left(\hat{\rho}(1),\hat{\mu}_1,\hat{\mu}_2,\hat{\mu}_3\right)=\left(0,181.8407,6.8710\times 10^4,3.8943\times 10^7\right).
\end{equation*}
Note the incapability of the estimate to capture the strictly positive lag-one autocorrelation coefficient. Then, the solution to (P2), where $\{\hat{D}^{(2)}_0(0),\hat{D}^{(2)}_1(0)\}$ is used as starting solution is given by
\begin{equation*}
\hD^{(2)}_0=\left(\begin{array}{cc}-0.0052 & 0.0052\\
0 & -1.3115\end{array}\right), \qquad \hD^{(2)}_1=\left(\begin{array}{cc} 0 & 0  \\
1.2240 & 0.0875 \end{array}\right).
\end{equation*}
with moments
\[\left(\rho_1,\mu_1,\mu_2,\mu_3\right) = \left(0,181.0040,6.99896\times 10^4,4.0497\times 10^7\right). \] The KL divergence is
\[D_{KL}\left(\{D_0,D_1\},\{\hD^{(2)}_0,\hD^{(2)}_1\}\right)=84.0645,\] clearly larger than (\ref{DKL_ex3}).  Finally, the log-likelihood function is
\begin{equation}\label{loglik_ex22}
\log f\left(\bt| \hD^{(2)}_0, \hD^{(2)}_1  \right)=-3051.7,
\end{equation}
which is smaller than $-3023.4$, therefore the estimate in first canonical form is selected.
}\Qed

\subsection{{Canonical versus redundant representation}}
The $\map_2$ can be expressed via either the redundant representation (\ref{known map}) or the canonical forms (\ref{can1}) or (\ref{can2}).  In principle, the only difference between the two representations is that the canonical one allows for a unique estimate of the model parameters, while the lack of identifiability of representation (\ref{known map}) implies possibly infinite estimates. However, the elements of interest associated with the $\map_2$, namely, the distributional properties of the variable $T$, are the same under equivalent representations. Also, if the interest is in the estimation of the $\map_2$$/G/1$ queueing system, \cite{Ramirez4} recently proved that the steady-state distributions coincide under equivalent arrival processes. Therefore, it is natural to wonder which are the benefits of using the estimates in canonical representation instead of the redundant ones.

\begin{figure}[htb]
\begin{center}
\begin{tabular}{c}
\includegraphics[height=2.7in]{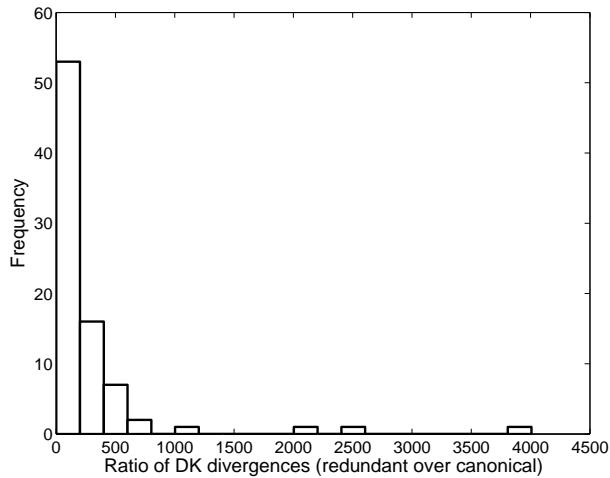}
\end{tabular}
\caption{\it Histogram of the ratio between the DK divergences of the redundant estimates over the DK divergences of the canonical ones.}\label{hist_can}
\end{center}
\end{figure}

To look more closely at this problem a hundred of random $\map_2$s
in redundant representation were simulated and estimated via a ML
approach equivalent to that described in Section \ref{eval_lik},
where the objective function is written in terms of the redundant
variables $\{\lambda_1,\lambda_2,p_{120},p_{110},p_{210},p_{211}\}$.
Here too the starting point was calculated as the solution of the
equivalent problem to (P0), where the moments are expressed in terms
of the 6 variables. Once the estimates were obtained, the DK
divergences between the real parameters and the estimated ones in
redundant version, were calculated. On the other hand, the canonical
estimates of the random $\map_2$s and their DK divergences were
computed using the ML method of Section \ref{eval_lik}. Figure
\ref{hist_can} depicts the histogram of the ratio between the DK
divergences of the redundant over the canonical estimates. It can be
seen that the DK divergence of the redundant forms are considerable
larger than those from the canonical versions and in consequence,
the canonical estimates are \emph{closer} to the true parameters
than the redundant ones.

It should be also pointed out that in eighteen out of the hundred of
simulated $\map_2$s, it was not possible to obtain the ML estimate
in redundant version. Apparently, the evaluation of the likelihood
function in terms of six parameters presents more numerical problems
than that in the canonical version, and in all these cases numerical
inconsistencies were found.

\section{Discussion}\label{sec: discussion}
In this paper we deepen our understanding of the maximum likelihood
estimation of the second-order $\map$, a suitable stochastic process
for many statistical modeling applications. Despite the apparent
straightforwardness of the problem, the matrix notation as well as
the intrinsic dependence structure of the process turn the
evaluation and maximization of the likelihood function into a
complicated task in practice. These difficulties are overcome by the
use of the canonical representation of the process, a proper
re-scaling of the objective function and a choice of a particular
starting solution of the algorithm. A method to compare between
different estimates is also delineated.

Prospects regarding this work may concern inference for higher order $\map$, which are expected to show more versatility for modeling purposes. We are aware of the complexity of such a problem due to the lack of unique representations and the increasing number of parameters. These complications present a challenging problem that we hope to address in the future.


In the spirit of a reproducible research the codes utilized in this paper to estimate the $\map_2$ are available at
\begin{center}
\noindent {\tt http://personal.us.es/jrcobo/www/Software.html}
\end{center}
as a stand-alone MATLAB${}^\copyright$ toolbox.

\section*{Acknowledgements}
Research partially supported by research grants and projects MTM2009-14039 (Ministerio de Ciencia e Innovaci\'on, Spain) and FQM329 (Junta de Andaluc\'{\i}a, Spain), both with EU ERDF funds.
The corresponding author is supported by Consolider "Ingenio Mathematica" through her post-doc contract.

\bibliographystyle{model5-names}
\bibliography{ref_paper}







\end{document}